\documentclass[footinbib,a4paper,aps,pra,reprint,twocolumn,preprintnumbers,amsmath,amssymb,10pt, superscriptaddress,english]{revtex4-2}
\usepackage{bbm}
\usepackage[english]{babel}
\usepackage{verbatim}
\usepackage{graphicx}
\usepackage{color}
\usepackage{tikz}
\usepackage[colorlinks=true,citecolor=blue,linkcolor=blue,urlcolor=blue]{hyperref}
\usepackage{braket}
\usepackage[normalem]{ulem}
\usepackage{upgreek}
\usepackage[percent]{overpic}
\usepackage{graphicx,xcolor}
\graphicspath{{figs/}}
\usepackage{dcolumn}
\usepackage{bm}
\usepackage[shortlabels]{enumitem}

\usepackage{bm}
\usepackage{physics}

\newcommand{\tc}[1]{\textcolor{blue}{#1}}
\newcommand{\new}[1]{{\color{red}{{#1}}}}
\newcommand{\old}[1]{\textcolor{green}{\sout{#1}}}

\usepackage[colorinlistoftodos,prependcaption,textsize=small]{todonotes}

\newcommand{\be}{\begin{equation}}
\newcommand{\ee}{\end{equation}}
\newcommand{\bfi}{\begin{figure}[htbp]\centering}
\newcommand{\efi}{\end{figure}}

\begin{document}

\title{Many-body localization regime for cavity induced long-range interacting models}
\author{Titas Chanda}
\affiliation{The Abdus Salam International Centre for Theoretical Physics (ICTP), Strada Costiera 11, 34151 Trieste,
Italy}
\author{Jakub Zakrzewski}
\email{jakub.zakrzewski@uj.edu.pl}
\affiliation{Institute of Theoretical Physics, Jagiellonian University, \L{}ojasiewicza 11, 30-348 Krak\'ow, Poland }
\affiliation{Mark Kac Complex
Systems Research Center, Jagiellonian University, \L{}ojasiewicza 11, 30-348 Krak\'ow,
Poland.}

\begin{abstract}
Many-body localization (MBL) features are studied here for a large spin chain model with long range interactions.
The model corresponds to cold atoms placed inside a cavity and driven by an external laser field with long range interactions coming from rescattering of cavity photons. Earlier studies were limited to small sizes amenable to exact diagonalization.
It is shown that nonergodic features and MBL may exist in this model for random disorder as well as in the presence of tilted potential on experimental time scales also for experimentally relevant system sizes using tensor networks algorithms.
\end{abstract}
\maketitle

\section{Introduction}

The
pioneering works \cite{Gornyi05,Basko06, Oganesyan07} started a modern area of intensive investigations of the interplay between the disorder and interactions. The belief that interacting closed systems  inherently locally thermalize
led to the eigenstate thermalization hypothesis (ETH) \cite{Deutsch91,Srednicki94}. {On the other hand,} the many-body localization (MBL) \cite{Gornyi05,Basko06, Oganesyan07} became a prominent example of ETH breaking\old{,} \new{--} various aspects of MBL are discussed in recent reviews \cite{Nandkishore15, Luitz17b, Alet18,Abanin19, Gopalakrishnan20}. {The} strong believe in the existence of MBL phase in the thermodynamic limit (even a proof of it was proposed for a certain class of spin chains \cite{Imbrie16,Imbrie16a})  was shattered by an influential recent contribution   \cite{Suntajs20e}, which was followed by a number of works providing arguments for and against the existence of MBL in the thermodynamic limit
\cite{Panda20,Sierant20b,Sierant20p,Abanin21,Suntajs20,Kiefer20,Sels20,Luitz20,Sierant21}. Similar conclusions could be obtained from approximate time dynamics for large system sizes \cite{Doggen18, Chanda20t,Chanda20m,Sierant21l}. In effect, recent claims do suggest the separation of the physical picture into the finite time, finite size, experimentally reachable ``MBL regime'' leaving the question of the existence of the ``MBL phase'' in the strict thermodynamic regime open \cite{Crowley21,Morningstar21,Sels21}.

\bfi
   \includegraphics[width=0.98\linewidth]{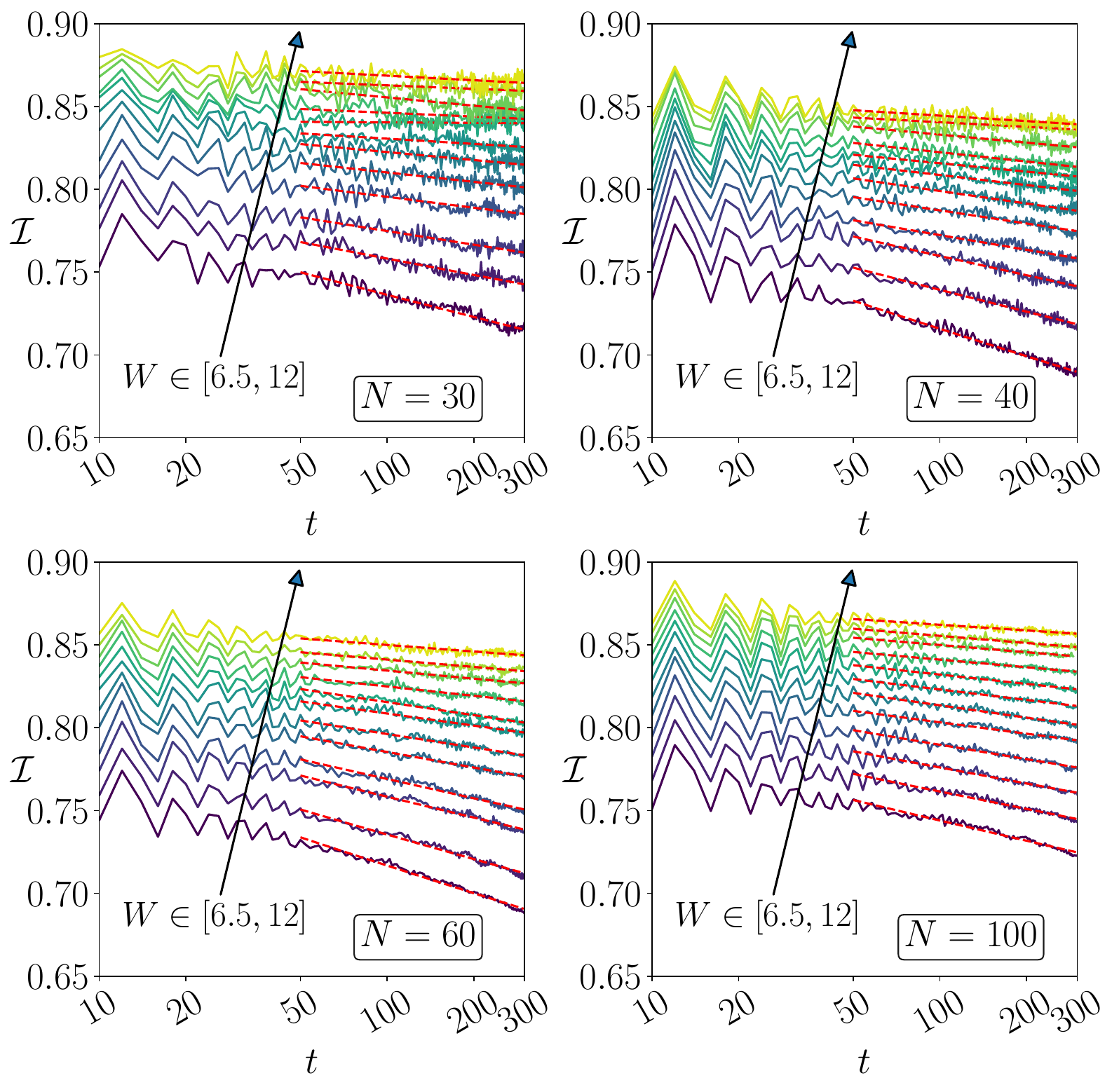}
  \caption{Imbalance as a function of time for different system sizes indicated in the panels and the all-to-all interaction strength $U_1=1$. The color codes 
  different disorder amplitudes $W$ (the higher $W$ the slower the decay -- highest lying curves correspond to the biggest $W$ {as indicated by the arrow pointing towards the direction of larger $W$s}). Dashed red lines yield power law, $t^{-\beta}$,  fits to the tails of the decay.}
  \label{figdis}
  \efi

These works concentrate mainly on 
spin$-1/2$ systems with nearest neighbor interactions. Experimental signatures of MBL {have been} mainly seen for spinful fermions in cold optical lattices \cite{Schreiber15,Choi16,Luschen17,Luschen18}.  Bosons {have} also {been} theoretically considered \cite{Sierant17,Sierant18} -- with a slight disadvantage {due to} a large effective local Hilbert space.

An early natural question has been posed whether MBL may be observed for long-range interactions. The pioneering experiment with effective Ising-type interactions {has been} realized in ion chain quantum simulator \cite{Smith16}. Several works {have} discussed the conditions on long-range tunnelings or interactions necessary for observing MBL, often concentrating on the relation between the power-law decay of the couplings $\sim 1/|i-j|^\alpha$ (with $i,j$ denoting lattice sites) and localization properties \cite{Yao14,Burin15,Burin15b,Maksymov17,MaksymovBurin20}. The lack of existence of MBL is postulated for $\alpha<2d$ {with $d$ being the dimension of the system} on the basis of perturbation type arguments.
{Actually, for isotropic interactions -- as is the case studied below -- the limit is much weaker, $\alpha +2 <2d$ \cite{Yao14,Burin15} for a discussion see \cite{MaksymovBurin20}.}

Surprizingly, recent works \cite{Sierant19c,Kubala21} have shown that signatures of MBL may be observed in a certain type of 
infinitely-ranged interacting Hamiltonians representing atoms interacting both via contact  and cavity-mediated interactions. The model considered consists of a one-dimensional (1D) atomic gas in an optical cavity and driven by an external laser field. Photon rescattering leads then to effective {infinitely-ranged} interactions {corresponding to the equality in $\alpha+2 =2d$ for one dimensional system.} While \cite{Sierant19c} 
considered both fermions and bosons MBL in such a set-up with an imposed random-diagonal disorder, one may consider also quasi-random interactions resulting from the mismatch between the wavelength of the external drive and the cavity \cite{Kubala21}. 


Both these studies {have} considered small system sizes amenable to the exact diagonalization. The question naturally arises whether MBL will exist also for
larger sizes, especially due {to being at the edge of the localization condition}
\cite{MaksymovBurin20}. Naturally, we cannot address the strict thermodynamic limit (and as seen from the discussion above, such a limit is controversial even for much simpler short-range interactions) but the modest question we want to address in this paper is whether the signatures of MBL
can be observed for experimental sizes and time scales as realized in cavity experiments \cite{Baumann10,Mottl11,Mottl12,Landig16,Hruby18}.
Contrary to {the} experiments, we shall limit ourselves to a 1D model \cite{Sierant19c} and use the state-of-the-art tensor network (TN) techniques \cite{Schollwoeck11,Orus14, Paeckel19} to study the time dynamics from appropriately prepared initial states. This is a long time established technique for ``experimental'' approach to MBL following the pioneering work \cite{Schreiber15}.

The paper is organized as follows. First we briefly recall the model and study the time dynamics in the presence of random disorder. Later we consider localization in disorder-free tilted lattice that extends extensive recent studies of such models \cite{Schulz19,vanNieuwenburg19,Taylor20,Chanda20c,Doggen20s,Yao20,Yao21,Yao21} to infinite range potentials. Let us also note here the recent work \cite{Li21} which discusses localization in tilted lattice for dipolar i.e., power-law decaying interactions.

\section{Model and methods}
\label{model}

{We} consider cold spin-polarized fermions in a 1D  optical lattice placed along a cavity axis. The cavity mode wavelength coincides with that
of the laser forming the optical lattice.
The optical lattice is sufficiently deep to facilitate a tight 
binding description of the system limited to its lowest band. {In addition atoms scatter  the laser light shined at them from the side.}  In the limit when cavity field may be adiabatically eliminated from the problem,
as studied often both experimentally and theoretically \cite{Larson08,Maschler08,Baumann10,Fernandez10,Habibian13,Landig16,Niederle16,Rojan16} (for a broad  reviews see \cite{Ritsch13,Mivehvar21}), the Hamiltonian of the problem reads 
(see \cite{Sierant19c} for a recent derivation):
\begin{align} \label{eq:hub}
     {H}_0  &=  \sum_{i=1}^{L-1}\left[{\frac{J}{2}}(\hat{f}_i^
    \dagger \hat{f}_{i+1} + \text{c.c.}) + V \hat{n}_i\hat{n}_{i+1}\right] \nonumber \\
    &-\frac{U_1}{L}\sum_{i\ne j }^{L}  (-1)^{i+j} \hat{n}_i\hat{n_j}
\end{align}
with $\hat{f}_i$ {being the}  annihilation operator for a fermion at site $i$, $\hat{n}_i=\hat{f}_i^\dagger\hat{f}_i$, $L$ is
the number of sites, $J$ describes the hopping between sites
and $V$ is the strength of the contact interactions. $U_1$ is the strength of cavity-mediated infinite range interactions, the scaling with the system size
in the last term above assures that the Hamiltonian \eqref{eq:hub}, is extensive in the large $L$ limit. 

The standard Jordan-Wigner transformation puts the Hamiltonian into Heisenberg spin chain formulation (for $J=V=1$  assumed {as the} energy unit later on) with infinite range interactions due to cavity induced term
\begin{equation} \label{eq:hei}
     {H}_0  =  \sum_{i=1}^{L-1}\vec{S}_i.\vec{S}_{i+1}  -\frac{U_1}{L}\sum_{{i\ne j }}^{{L}}  (-1)^{i+j} {S}^z_i{S^z_j}
\end{equation}
where $\vec{S}_i$ is 1/2-spin at site $i$. 
The system is modified  by an additional random diagonal disorder term $H_r=\sum_{i=1}^L h_i \hat{n}_i$ (corresponding to $H_r=\sum_{i=1}^L h_i S^z_i$ in the spin language) where $h_i$ are the random variables with uniform distribution in $[-W,W]$ interval. The full Hamiltonian becomes:
\be \label{eq:mod}  
H=H_0+H_r. 
\ee
We shall also consider the disorder free situation in Sec.~\ref{free} where $h_i$ becomes a simple function of the position. Specifically we consider a tilted lattice, $h_i=Fi$ with $F$ being the magnitude of the tilt, or \tc{we} assume that $h_i$ comes from the harmonic trapping potential.  We consider a half filling case with $L/2$ fermions (which corresponds to the largest spin sector $\sum S^z_i=0$) and consider time evolution of the initial charge-density wave Fock-like separable states. However, contrary to standard MBL approaches for short range interactions where, following experiments \cite{Schreiber15,Luschen17}, the N\'eel state $\ket{\psi(0)}= |1,0,1,0,...\rangle$ ($|\uparrow,\downarrow,\uparrow.\downarrow,...\rangle$ in the spin language) is often assumed, we need a different strategy. For short range interactions and random disorder the model exhibits, in finite size studies, a mobility edge \cite{Luitz15}, the state $\ket{\psi(0)}$ being most difficult to localize, making it a good candidate as a probe of dynamics. This is no longer true in the presence of all-to-all interactions with non-negligible $U_1$ -- the $(-1)^{i+j}$  factor in  the last term of \eqref{eq:hei}
shifts the N\'eel state either close to the bottom or to the top (depending on the sign of $U_1§$) of the spectrum. Thus, to probe the system, we consider a sample of random initial separable Fock-like states that we choose to lay in the middle of the spectrum. For each realization of the disorder we find states of minimal and maximal energy by density matrix renormalization group (DMRG) algorithm \cite{White92,White93} and consider random states with energy close to the middle energy {(see \cite{Chanda20m} for the details about the description of choosing the initial state near a given energy}).

The numerical study of time dynamics is carried out using Time Dependent Variational Principle (TDVP) {in the} tensor network approach \cite{Haegeman11,Haegeman16,Koffel12,Paeckel19}, {with an} implementation based on the Itensor library \cite{itensor} that has been thoroughly tested in out earlier works on short-ranged interacting systems \cite{Chanda20,Sierant21}. The diagonal all-to-all interaction term is efficiently represented as a matrix-product-operator enabling studies of realistic system sizes for times corresponding to typical times in cold-atoms time-dynamics experiments \cite{Scherg21}. The time dynamics is addressed following the time evolution of the correlation function
\be
{\cal I}(t) =\frac{4}{L} \sum_{i=1}^L \langle S^z_i(t)\rangle \langle S^z_i(0)\rangle 
\label{imb}
\ee
which for a N\'eel state would reduce to a standard imbalance of occupations between odd and even states.

\section{The disordered case}

Let us consider first a random uniform disorder. Fig.~\ref{figdis} shows the characteristic time dynamics of the ``imbalance''  or magnetization {correlation function} \eqref{imb} for different system sizes $L$ and for a non-negligible all-to-all interaction strength $U_1=1$. 
 After a rapid initial decay from ${\cal I}(0)=1$ on a time scale of few hopping times (not shown) the long time dynamics is clearly dependent on the disorder amplitude $W$.  The red dashed lines  are approximate power-law ${\cal I}(t)\sim t^{-\beta}$ fits that work remarkably well. That resembles similar observations  for  short range interacting models \cite{Luitz16, Luschen17,Chanda20t,Chanda20m}. On the other hand, comparing the disorder amplitudes 
 for which data are obtained, a careful reader will notice that the slow-down of the dynamics and possible saturation of the imbalance occurs for
 much stronger disorder than for corresponding disordered Heisenberg spin chain \cite{Doggen18,Chanda20t} where such a saturation is observed around $W\approx 4-5$.
 
 {We must stress that a determination of the disorder amplitude which may be considered as a border value for localization is notoriously difficult~\cite{Luitz16,Doggen18,Chanda20t,Sierant21}. For small system sizes, the algebraic decay observed after a rapid initial decrease of imbalance eventually slows down {at} long times {--} the imbalance reaches a final non-zero value (for a sufficiently large disorder) well beyond the Heisenberg time \cite{Sierant21}. Such long times are unreachable for larger system sizes as Heisenberg time increases exponentially with the system size. Moreover, the current experimental capabilities limit the coherent, decoherence-free evolution in cold atoms to few hundreds of the tunneling time.
 {Therefore, as the study of  the evolution gets limited in time,}
 one is forced to define {a} critical {value of the power-law exponent} $\beta_c$ {below} which we may consider the system to be  localized. One of the possibilities {to get a correct estimate of $\beta_c$} is to again refer to small systems and define $\beta_c$ {equal to the power-law exponent that occurs} for such a disorder for which the level {statistics} is already close to be Poissonian indicating localization \cite{Chanda20t}. This leads to $\beta_c\approx 0.02$ for short-ranged model. The other possible way is to
 adopt the criterion that vanishing $\beta$ within error bars -- coming from averaging over the disorder realizations  -- is sufficient \cite{Doggen18} {so that $\beta_c$ becomes comparable with the standard deviation $\sigma$ of $\beta$.}
 For a sample of few hundreds of disorder realizations for $L=16$ as in \cite{Doggen18}, this leads, taking $\beta_c=\sigma$, to $\beta_c \approx 0.01$ for short-ranged model. We adopt  the same criterion {of $\beta_c = 0.01$} for simplicity.

\begin{figure}[t]
   \includegraphics[width=0.6\linewidth]{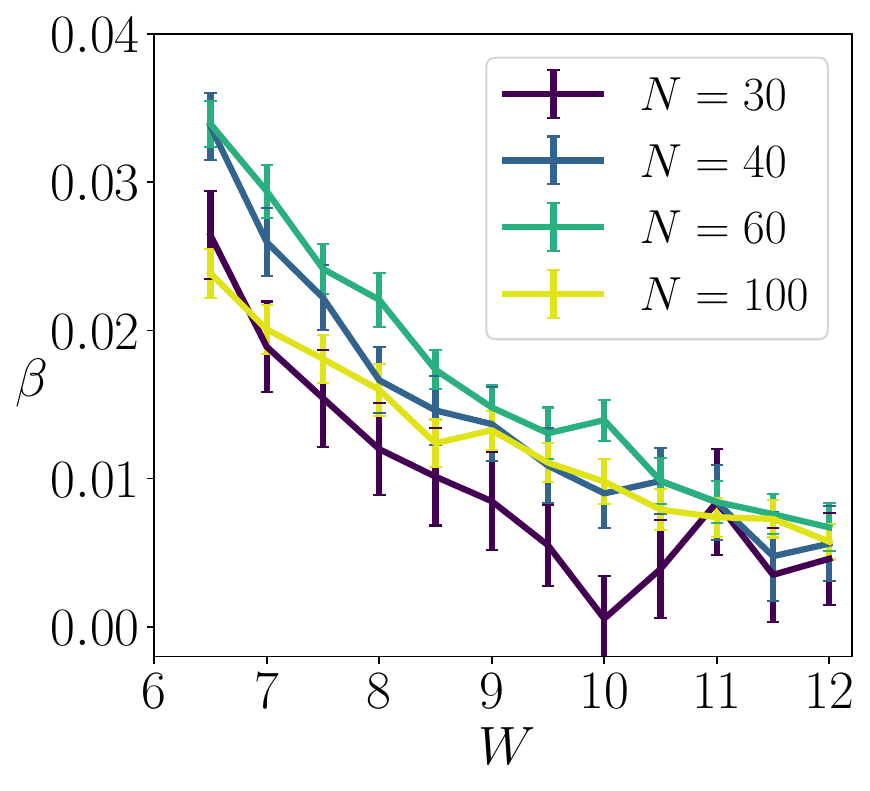}
  \caption{The average power-law decay of the imbalance for different system sizes. 
{Here we plot the power-law exponent $\beta$ as a function of the disorder strength $W$ for different system sizes.}  
  The critical $\beta=0.01$ is reached around $W=10.5$.}
  \label{figdis40}
\end{figure}
 
 \begin{figure*} [ht] \centering
   \includegraphics[width=0.8\linewidth]{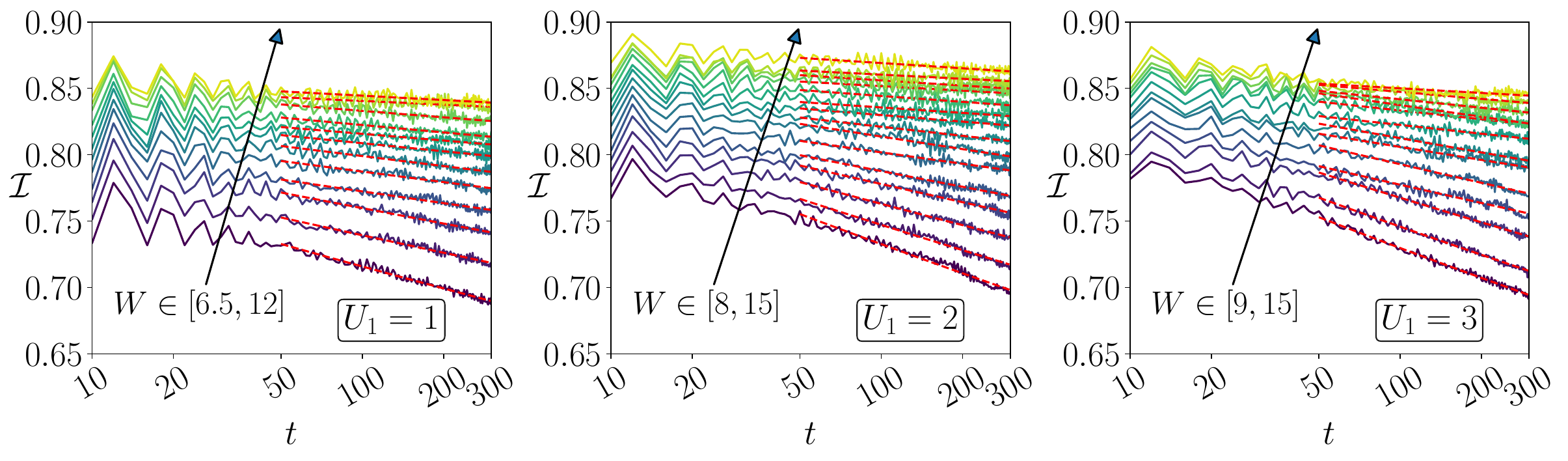}
  \caption{Imbalance as a function of time for different all-to-all interaction strength $U_1$ and different, color coded disorder amplitudes $W$ (as in Fig.~\ref{figdis}) for the system size $L=40$. Dashed red lines indicate power law fits to the data. }
  \label{figdisU}
  \end{figure*}
 
 To {support this approach further, we} provide a more quantitative comparison of the saturation {by fitting}  the  the power-law to the data for different system sizes {as summarised in}  Fig.~\ref{figdis40}.
  While some system size dependence is still observed despite reaching $L=100$, $\beta$ clearly decays with $W$. 
  {The adopted critical $\beta_c=0.01$ value is reached around {$W \approx 10.5$}.} We note that the obtained value of the critical $W$ estimated in this way quite nicely corresponds to
 that obtained via a finite size scaling of exact diagonalization results \cite{Sierant19c}.

 \bfi
   \includegraphics[width=0.6\linewidth]{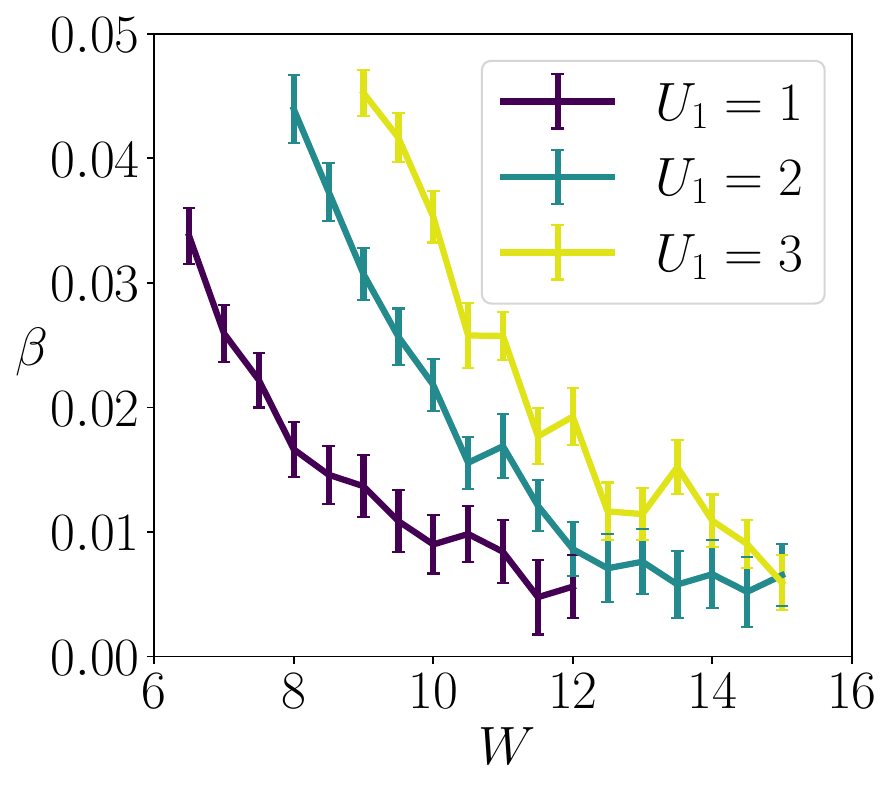}
  \caption{Fitted values of $\beta$ for $L=40$ and different $U_1$ values. The localization crossover shifts to larger disorder for stronger all-to-all interactions $U_1$.}
  \label{betaU}
  \efi

 The work \cite{Sierant19c} have shown that the localization crossover point is strongly dependent on $U_1$.  To see whether this conclusion persists for larger system sizes we fix the system size at $L=40$ and inspect the time dynamics for few values of $U_1$, c.f., Fig.~\ref{figdisU}. We again observe that the long time dynamics {(here we extend our studies to 300 tunneling times, the value reachable with our algorithm for $L=40$)} is well approximated by power-law behavior {for different values of $U_1$}. Then we can estimate the crossover to a localized model for different $U_1$, see Fig.~\ref{betaU}. We observe that the stronger $U_1$ (in the small interval studied) the bigger $W$ are needed to reach the localization border, again in good agreement with results of \cite{Sierant19c} obtained for much smaller system sizes.

 At this point let us comment that our numerical results  does not provide any information about the thermodynamic limit in the problem studied. Even for short-range interacting systems this remains an open question (see e.g., \cite{Suntajs20,Panda20,Sierant20,Laflorencie20,Abanin21,Sels21,Sierant20p,Sierant21}). We merely show that for realistic system sizes and  experimentally reachable evolution times the dynamics of the Hamiltonian \eqref{eq:hub} is strongly nonergodic and, for sufficiently strong $W$, shows localized behavior.
  
\section{Disorder-free models}
\label{free}

Consider now the second, already announced in Sec.~\ref{model}  choice of $h_i$ in $H_r$ which becomes a smooth function of the site index. To be specific consider first  $h_i=Fi$ where $F$ is a constant tilt of the lattice realized, e.g., \cite{Scherg21} by the Zeeman effect. Without the optical cavity the model \eqref{eq:mod} reveals the so called Stark many-body localization
(SMBL) which has been intensively studied since its discovery for spinless fermions \cite{vanNieuwenburg19,Schulz19} as well as for bosons \cite{Taylor20,Yao20b,Yao21}, the experimental verification came with spinful fermions in optical lattice \cite{Scherg21} as well as in quantum simulators \cite{Guardado20,Guo20,Morong21}. It has been also studied for open systems \cite{Wu19}. Here we briefly consider whether the presence of the long-range cavity induced interactions affects SMBL.

For the Heisenberg spin chain the transition to SMBL is observed in dynamical studies for $F$ of the order of unity. Here we consider larger values
of $F$ and show that the short time dynamics is characterized by the time scale inversely proportional to $F^2$, c.f., Fig.~\ref{figU1f}, in a similar manner to that observed for short-range Hamiltonians \cite{Yao21} and explained there by a pair tunneling mechanism that preserves the approximate global dipole moment, ${\cal D} = \sum_i (i-i_0) f^\dagger_if_i$. {Note that, for convenience, we evaluate the dipole moment with respect to the center of the system, where $i_0$ is half-integer for even number of sites considered.} {The dipole moment preserving}   second order process corresponds to a tunneling of one particle by one site to the left while the other by one site to the right  and is given by an effective tunneling rate $\tilde J = 2VJ^2/F^2$ {in the absence of all-to-all interactions} \cite{Yao21} (for a detailed discussion of processes occurring at large $F$ values see \cite{Taylor20}). {The presence of all-to-all interactions affects slightly this picture. The second order pair tunneling process has now contributions with slightly different energies: 
\begin{align}
\tilde J &= \frac{J^2}{-F-V-k_1U_1/L} +  \frac{J^2}
{F-V-k_2U_1/L}\nonumber \\
&\approx \frac{J^2[2V+(k_1+k_2)U_1/L]}{F^2},
\label{eq:new}
\end{align}
where $k_1$ and $k_2$ are small positive or negative integers giving the energy difference due to all-to-all interactions for the corresponding
single particle jump. Since $U_1/L\ll V$ the additional terms proportional to $k_i$  may be to a large extent neglected -- see, however, below.} 

\bfi
   \includegraphics[width=0.98\linewidth]{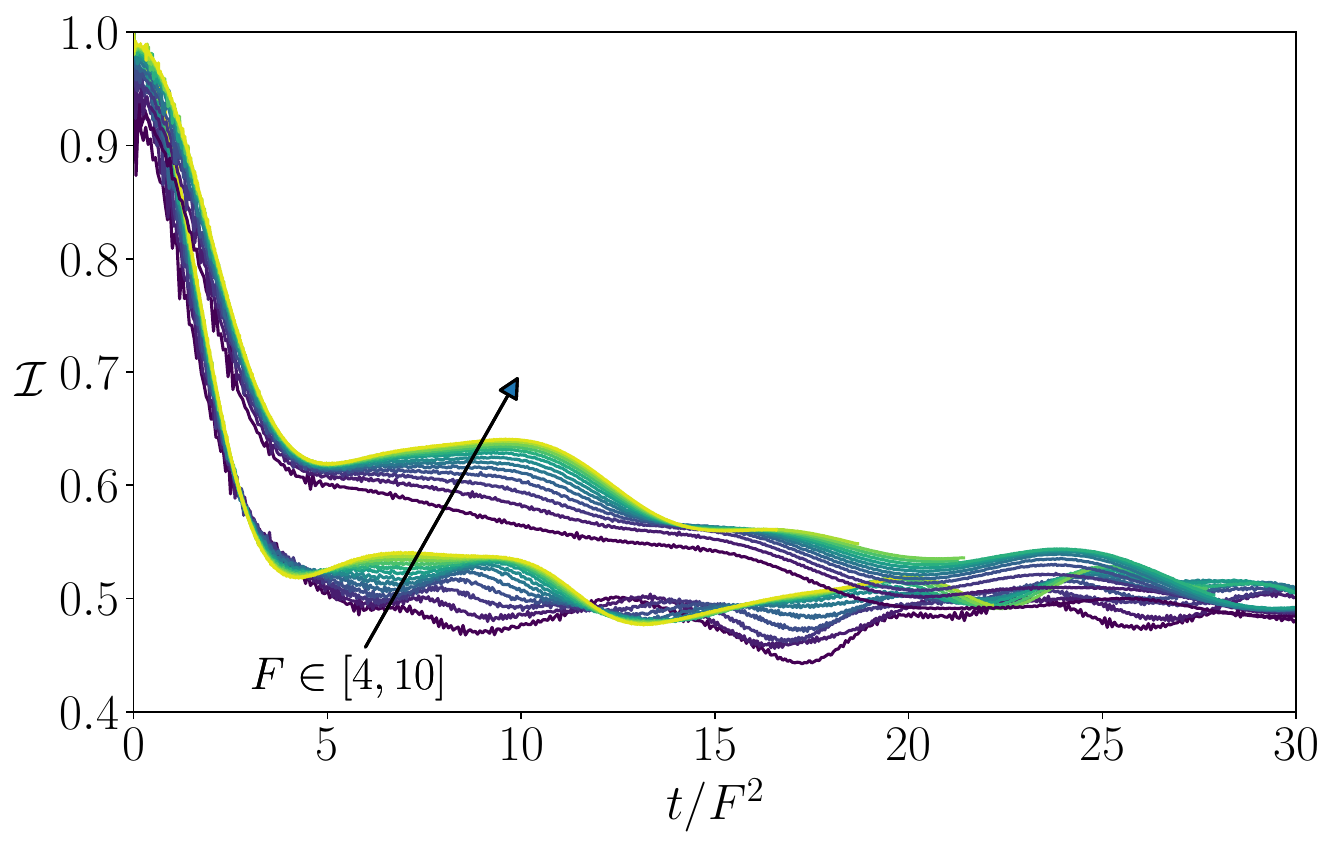}
  \caption{The time dynamics of the imbalance for the tilted lattice for $U_1=1$ for smaller system $L=20$ (lower curves) and $L=40$ (upper curves). Different colors correspond to different values of $F\in[4,10]$ from dark blue to yellow lines for $F=10$. The initial slow decay of imbalance 
  is ruled by the effective second order tunneling (see text) with time scale proportional to $F^2$. This scaling breaks for longer times due to cavity mediated interactions.}
  \label{figU1f}
\efi  

To obtain time dynamics of the imbalance  (in the absence of disorder) we consider an average over 100 initial separable states. They are not chosen, as in the previous section, by the energy condition but rather we impose that all the initial states have the same dipole moment ${\cal D}=0$. Choosing a vanishing dipole {moment} corresponds roughly to the middle {of the} energy in Hilbert space shattered \cite{Khemani20} spectrum. By  taking the same dipole moment we ensure that the initial states belong to the same ``sector'' in the dynamics.
The data in Fig.~\ref{figU1f} are presented for a single value of all-to-all interactions $U_1=1$ but for two system sizes $L=20$ and $L=40$. $L=20$ curves could be also obtained by a  standard Chebyshev propagation \cite{Fehske08}, we present them as a comparison to a typical $L=40$ case since tilted lattices with cavity mediated infinite range interactions have not been analyzed before. 

For both system sizes the results are quite similar.
 The initial decay, apart from $F^2$ scaling, shows some dependence on the system size -- initially the decay is slower for the bigger system, while at longer time imbalance for the smaller system saturates faster. This may be attributed probably to the hyper-extensive character of the tilted part of the Hamiltonian -- thus for larger systems the
approximate conservation of the global dipole shows more strongly {-- as apparent also from \eqref{eq:new}}. Their long time dynamics, apart from an apparent saturation for significantly non-zero values of the imbalance (revealing localization) is characterized by long period oscillations. Again a careful comparison of both sets of curves reveals that this period reflects the system size. The oscillations, absent for short-ranged Hamiltonian are related to cavity mediated interactions that are system size dependent via $U_1/L$ term. {We believe that they may be related to equally spaced (by integer multiples of $U_1/L$) corrections to energies appearing above  in the  second order effective tunnelings.}

{The most important effect of the all-to-all interactions is a significant reduction of the final imbalance values. For short-ranged interactions with $V=1$ those values were quite well reproduced by interaction-free analytic expression 
\begin{equation}
I_{\rm fin}={\cal J}_0^2(\frac{2J}{F}),
\label{eq:bes}
\end{equation}
where ${\cal J}_0(x)$ is the {zeroth-order} Bessel function thus being above 0.85 for $F\ge4$. For $U_1=1$ the observed imbalance is twice lower indicating that all-to-all interactions play non-trivial role in the dynamics.} 

\bfi
   \includegraphics[width=0.98\linewidth]{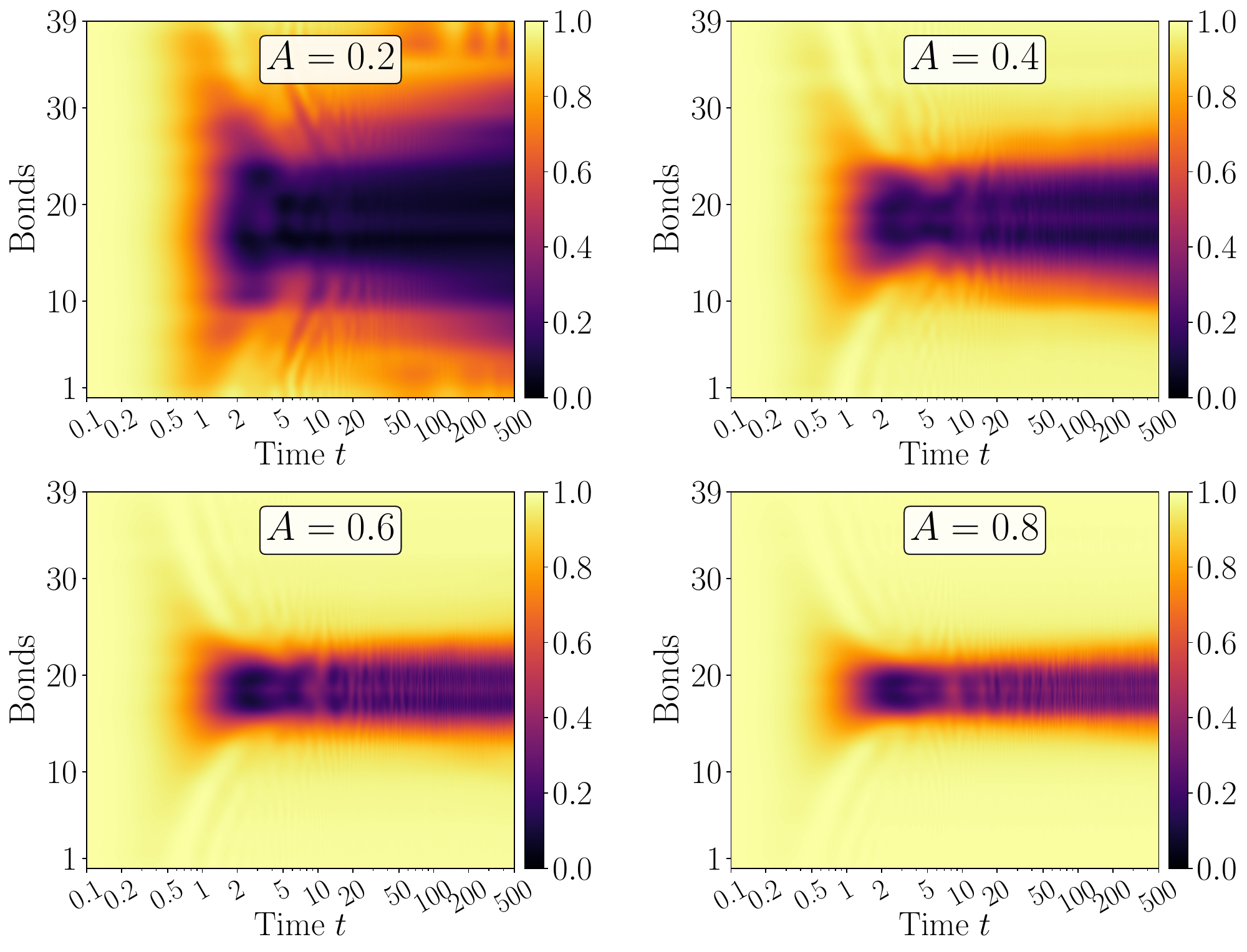}
  \caption{Time dynamics of the local imbalance, \eqref{locI} for different curvatures, $A$, of the harmonic potential placed on top of the lattice.
  Observe the coexistence of the delocalized central region with outer parts where local imbalance is practically unaffected by dynamics revealing very strong localization.}
  \label{figU1loc}
\efi  

As a second example we consider the case of a harmonic potential on top of the lattice, $h_i=\frac{A}{2}(i-i_0)^2$, where $i_0$ denotes the center of the lattice ($i_0=20.5$ for lattice of size $L=40$ considered). Such a potential {has been} considered in \cite{Chanda20c} where we have found that for 
short-ranged interactions the system splits into separate regions. While at the center the system delocalizes, clear signatures of localization are observed at the outer regions leading to ``phase separation'' or the coexistence of localized phases with the ergodic central bubble. This has been attributed to the fact that the local dynamics is governed by a local static field, $F_{loc} = A(i-i_0)$, following the first order Taylor expansion of
the potential. Indeed soon such a coexistence phenomenon was verified in the quantum simulator experiment \cite{Morong21} where the system may be described by the long-range-interacting Ising chain. 

Figure~\ref{figU1loc} shows that this is the case also for cavity induced interaction system, \eqref{eq:mod}. To visualize local dynamics we use the concept of local imbalance introduced by us previously \cite{Chanda20c,Yao21}. In the spin formulation  {the local imbalance} reads
\begin{equation}
I(i)=2 |\langle S^z_i0)\rangle\langle S^z_i(t)\rangle + \langle S^z_{i+1}(0) \rangle\langle S^z_{i+1}(t)\rangle|,
\label{locI}
\end{equation} 
while for spinless fermions picture one should replace $S^z_i$ by $f^\dagger_if_i -1/2$. To be precise $I(i)$ gives the local imbalance between sites $i$ and $i+1$ so it is defined on the bond $i$ linking these two neighboring sites. As we may observe in Fig.~\ref{figU1loc} the extend of the dark delocalized region shrinks with the increasing curvature $A$ of the potential. A careful inspection of the time dynamics reveals the very slow growth
of the central region in time even for large $A$. This could be mapped to a small non-zero $\beta$ observed for disordered cases even for large disorder amplitudes and with the slow decrease of the imbalance hidden under pronounced oscillations in Fig.~\ref{figU1f}. This phenomenon 
may indicate possible delocalization of the system for very long times. Here we limit the evolution to experimentally feasible times of the order of $t=500/J$~\cite{Scherg21}.

\bfi
   \includegraphics[width=0.6\linewidth]{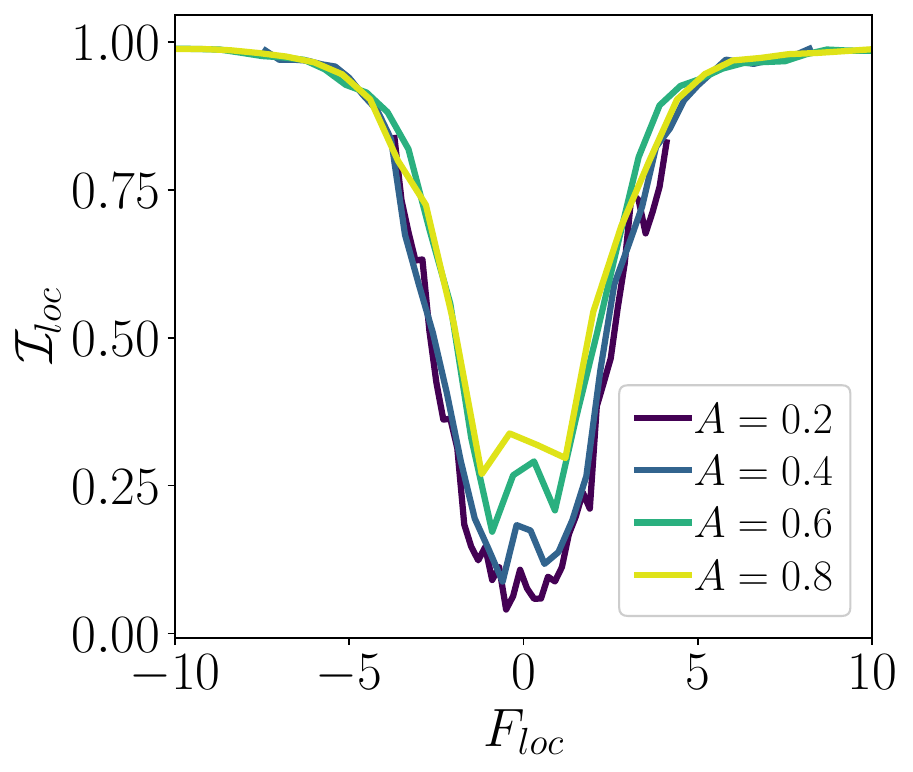}
  \caption{
 {The long time snapshot of the local imbalance as a function of the local effective field value $F_{loc}$  for the harmonic perturbation of the lattice. Here we consider a system of size $L=40$ and different values of the harmonic curvature $A$. The long time snapshot has been computed by averaging the date over the time window $t \in [400/J, 500/J]$. }
}
  \label{figU1ss}
\efi  

{Furthermore, earlier works \cite{Chanda20c,Yao21} hinted that for short-ranged interacting systems the transition in space between the localized and delocalized parts of the system depends only on the local effective field value $F_{loc} = A(i-i_0)$ irrespective of the curvature value $A$. To verify this for cavity induced infinite-ranged system, we plot the variations of the long time snapshot of the local imbalance as a function of the local effective field value $F_{loc}$ in Fig.~\ref{figU1ss}. The collapse of the curves for different values of $A$ strongly suggests that this is indeed the case also for the infinite-ranged interacting system.}

{\section{Discussion}
While MBL is typically considered for short-range interacting systems, the present work shows that strong signatures of non-ergodic dynamics
occur for systems with infinite range (all-to-all) interactions. {T}his has been observed for small system sizes \cite{Sierant19c,Kubala21} amenable to exact localization, we confirm that the similar behavior characterizes larger, experimentally relevant systems. The question arises to what extend this non-ergodic behavior is related to a specific form of all-to-all density-density interactions (compare \eqref{eq:hub} or \eqref{eq:hei})
in which no coherent long-range coupling (e.g., long range tunneling) is present. 
{We} {r}ecall also that  the interactions considered are at the border of the condition derived for uniform long range interactions \cite{Yao14,Burin15,MaksymovBurin20} discussed in the Introduction. On the other hand, recent studies of lattice gauge theories indicated \cite{Chanda20} that 
the confinement and the lack of thermalization may occur due to global constrains in the system as those imposed by gauge theories (for recent related reviews see \cite{Banuls20,Aidelsburger22}.}

We have studied the disordered model
 of spinless fermions interacting via contact interactions as well as cavity mediated all-to-all forces extending our previous studies to large system sizes. For the standard diagonal disorder we observe that for realistic, experimentally relevant system sizes as well as experimentally achievable coherent evolution times a crossover from delocalized to localized phase persists for large system sizes. Moreover
the corresponding disorder amplitude nicely matches finite size scaling prediction \cite{Sierant19c}. 

We have also analyzed localization properties for the all-to-all cavity mediated interactions for disorder-free potentials showing that for moderate $U_1$ both Stark many-body localization and phase coexistence can be observed both for small and larger, of the order of 100 sites systems. 
{We have considered strongly tilted systems where strong Stark MBL is observed in the absence of infinite range interactions and the Hilbert space has a characteristic shattered character due to approximate conservation of the global dipole moment. Due to large oscillations present in the correlation functions the results were averaged over several initial states corresponding to the same value of the global dipole moment.  
We must} stress again that our conclusions are limited to finite system sizes and finite times corresponding to several hundreds of the tunneling times. 
Such time scales are relevant for experiments where the coherence times reach at most 700-1000 tunneling times. We observe indications that
the systems studied may very slowly delocalize on a much larger time scale but this phenomenon is not possible to be addressed by current tensor network algorithms. 

We have considered moderate $U_1$ values. One may consider what happens if $U_1$ term dominates the Hamiltonian. This term is diagonal
in the Fock site basis and as such is expected to lead on its own, in the limit of $U_1\rightarrow\infty$ to Hilbert space fragmentation into parts corresponding to different values of the almost conserved quantity
 \be {\cal K}=\sum_{i<j}  (-1)^{i+j} {S}^z_i{S^z_j}=\left(\sum_i (-1)^i S^z_i\right)^2 - constant.
 \ee
 That may lead to additional nonergodic features of the dynamics in this regime, in an analogy to large $F$ limit.

\section*{Acknowledgments}

The numerical effort was possible thanks to the support of PL Grid infrastructure. 
The support by National Science Centre (Poland) under project OPUS 2019/35/B/ST2/00034 (J.Z.) is acknowledged.

%

\end{document}